\documentclass[aps,prl,preprint,superscriptaddress,showpacs]{revtex4}

\usepackage{latexsym}
\usepackage{graphicx}  
\usepackage{epsfig}    
\usepackage{dcolumn}   
\usepackage{bm}        

\newcommand{\vk}{\mathbf{k}}

\begin{document}

\title{Coherent forward stimulated Brillouin scattering of a spatially incoherent laser beam in a plasma and its effect on beam spray}

\author{M. Grech}
\email[mickael.grech@gmail.com]{ }
\affiliation{Max Planck Institute for the Physics of Complex Systems, 01187 Dresden, Germany}
\affiliation{Centre Lasers Intenses et Applications, Universit\'{e} Bordeaux 1 - CNRS - CEA, 33405 Talence
Cedex, France}

\author{G. Riazuelo}
\affiliation{D\'{e}partement de Physique Th\'{e}orique et Appliqu\'{e}e,
CEA/DAM Ile-de-France, 91297 Arpajon Cedex, France}

\author{D. Pesme}
\affiliation{Centre de Physique Th\'{e}orique, Ecole Polytechnique, 91128 Palaiseau Cedex, France}

\author{S. Weber}
\affiliation{Centre Lasers Intenses et Applications, Universit\'{e} Bordeaux 1 - CNRS - CEA, 33405 Talence
Cedex, France}

\author{V. T. Tikhonchuk}
\affiliation{Centre Lasers Intenses et Applications, Universit\'{e} Bordeaux 1 - CNRS - CEA, 33405 Talence
Cedex, France}

\date{\today}

\begin{abstract}
A statistical model for forward stimulated Brillouin scattering (FSBS) is developed for a spatially incoherent, monochromatic, laser beam propagating in a plasma. A threshold for the average power in a speckle is found, well below the self-focusing one, above which the laser beam spatial incoherence can not prevent the coherent growth of FSBS. Three-dimensional simulations confirm its existence and reveal the onset of beam spray above it. From these results, we propose a new figure of merit for the control of the propagation through a plasma of a spatially incoherent laser beam.
\end{abstract}

\pacs{52.35.Mw, 52.35.Fp, 52.38.Hb, 52.38.Bv, 42.65.Jx}
\maketitle


Controlling the propagation of randomized laser beams through large scale plasmas is a crucial issue for inertial confinement fusion (ICF)~\cite{lindl_POP_95}. Coupling of the high intensity laser electromagnetic wave to ion and electron plasma waves can lead to backscattering, spreading or bending of the laser light~\cite{insta_para}. To avoid such deleterious effects and to improve the laser illumination uniformity, laser beam smoothing techniques have been designed~\cite{kato_lehmberg}. The resulting laser intensity distribution in the focal volume is made of many randomly distributed in time and/or in space spikes, the so-called speckles. The statistical properties of such smoothed laser beams are known in vacuum~\cite{rose_garnier}, but can be strongly modified by the interaction with the plasma. This phenomenon, referred to as plasma induced smoothing, arises from forward scattering of the laser light on the laser induced density perturbations in the plasma~\cite{lissage_plasma1,maximov_POP_01,lissage_plasma2}. Several processes of forward scattering on such perturbations, namely, self-focusing (SF), filament instability (FI), forward stimulated Brillouin scattering (FSBS) and multiple scattering (MS) have been considered.

The relative importance of each of these processes depends on the average power carried in a speckle, $\langle P\rangle$, compared to the critical power for SF~\cite{insta_para}: $P_c = (8\pi/k_0^2)\,c\,n_c\,T_e\,n_c/n_0$, where $k_0=2\pi/\lambda_0$ is the laser wave number, $c$ is the light velocity, $n_c$ is the critical density and $n_0$ and $T_e$ are the electron density and temperature, respectively. When $\langle P\rangle$ is about or above $P_c$, most of the speckles in the focal spot vehiculate a power above $P_c$ and are thus unstable with regard to SF and FI. This regime corresponds to strong beam spray~\cite{lissage_plasma1,maximov_POP_01}.

At lower intensities, plasma induced smoothing is due to FSBS seeded by the laser light MS on plasma density perturbations~\cite{lissage_plasma2}. FSBS is well described in the case of a coherent incident pump wave~\cite{eliseev_POP_97}. The case of FSBS driven by spectrally broadened pump has been investigated in Refs.~\onlinecite{lushnikov} in the limit of very short coherence times. It is shown that, while SF is prevented by the pump incoherence, FSBS can be the dominant instability. In Ref.~\onlinecite{maximov_POP_01}, the authors considered the case of a spatially incoherent, monochromatic, pump. They showed that the spatial growth rate of light scattered at large angles (compared to the incident beam aperture) is strongly reduced compared to the coherent case. However they restricted their study to the evolution of the average amplitude of the scattered electromagnetic wave . We will show that a correct description of the instability actually follows from the study of both daughter waves.


In this Letter, a statistical model for FSBS driven by a monochromatic, spatially incoherent, pump wave is developed. Propagation of the laser beam in the underdense plasma with average electron density $n_0 \ll n_c$ is described within the paraxial approximation. The wave electric field is separated into two components, the fields $a^{(p)}_{\vk}$ and $a^{(d)}_{\vk}$, associated with the incident laser beam and scattered wave, respectively. The vectors $\vk$ are lying in the plane perpendicular to the beam propagation axis $z$. We assume small scattering angle, $\theta_d \ll 1$. For a given scattering angle $\theta_d$, we  denote as $\vk_d$ the transverse wave number of the scattered wave $\vk_d = k_0\, \theta_d$, and $\vk_s = - \vk_d$ the ion acoustic wave wave-number associated to this scattered wave. The frequencies of the pump and scattered wave are $\omega_0$ and $\omega_0 - \Omega^{(s)}_{\vk_s}$, respectively, with $\Omega^{(s)}_{\vk_s}= c_s\vert\vk_s\vert$. The equation for the scattered wave reads:
\begin{eqnarray}
\label{eq1a} \big(c\,\partial_z+i\,\Omega^{(d)}_{\vk}\big)\,a^{(d)}_{\vk} &=& \gamma_0\,\int d\vk_p\,a^{(p)}_{\vk_p} a^{(s)*}_{\vk_p-\vk}\,,
\end{eqnarray}
where $\Omega^{(d)}_{\vk}=c^2\vk^2/2\omega_0$ is the frequency detuning due to diffraction~\cite{remark}. The pump and scattered electromagnetic waves are coupled to the ion acoustic wave $a^{(s)}_{\vk}$ with frequency $\Omega^{(s)}_{\vk}$. It propagates with the velocity $c_s\ll c$ in the transverse plane. This low frequency plasma response to the ponderomotive force is described by a time-enveloped wave equation:
\begin{eqnarray}\label{eq1b} 
\big(\partial_t+\nu_s+i\,\Omega^{(s)}_{\vk}\big)\,a^{(s)}_{\vk} =\gamma_0\,\int d\vk_p\,a^{(p)}_{\vk_p} a^{(d)*}_{\vk_p-\vk} \,,
\end{eqnarray}
where $\nu_s \ll \Omega^{(s)}_{\vk_s}$ is the ion acoustic damping rate. Amplitudes $a_{\vk}^{(p,d,s)}$ are dimensionless and the FSBS coupling constant is $\gamma_0^2 = \Omega^{(s)}_{\vk_s}\,\omega_0\,(n_0/n_c)\,\langle I \rangle/(8c\,n_c\,T_e)$, where $\langle I \rangle$ is the incident laser average intensity.

Statistical properties of the spatially incoherent, monochromatic, pump can be prescribed. For usual spatial smoothing technics, it follows Gaussian statistics with a zero mean value~\cite{rose_garnier}:
\begin{eqnarray}
\label{eq2c}\langle a^{(p)}_{\vk_p} \rangle = \langle a^{(p)}_{\vk_p}\,a^{(p)}_{\vk_p'} \rangle =0\,,\,\, \langle a^{(p)}_{\vk_p}\,a^{(p)*}_{\vk_p'} \rangle = n^{(p)}_{\vk_p}\,\delta(\vk_p-\vk_p'),\,\,
\end{eqnarray}
where $n^{(p)}_{\vk_p}$ is the pump spatial spectrum normalized to one: $\int d\vk_p\,n^{(p)}_{\vk_p}=1$. It does not depend on the longitudinal coordinate $z$ as long as the propagation distance is shorter than the beam Rayleigh length (which is typically a few mm under ICF conditions). For applications, a Gaussian spectrum is considered: $n^{(p)}_{\vk_p}=(\rho_0^2/2\pi)\,\exp\!\big(\!-\!\rho_0^2\,\vk_p^2/2\big)$,  where $\rho_0$ is the speckle width.

Due to the pump incoherence, the scattered and acoustic fields are stochastic quantities and their statistical proprieties must be adressed. This is done by using the iterative method developped in Ref.~\onlinecite{thomson}, which allows to calculate the successive momenta of $a^{(d,s)}_{\vk}$. In this Letter, we restrict ourselves to study of the average fields $\langle a^{(d,s)}_{\vk} \rangle$ and take, as a criterium for the instability, the condition that at least one of the average field $\langle a^{(d)}_{\vk_d} \rangle$ or $\langle a^{(s)}_{\vk_s} \rangle$ increases in time and/or space. We do not compute the average intensities $\langle\vert a^{(d,s)}_{\vk}\vert^2\rangle$ because they can only evolve faster than ${\rm max}\big\{\vert\langle a^{(d)}_{\vk}\rangle\vert^2,\vert\langle a^{(s)}_{\vk}\rangle\vert^2\big\}$. Applying the iterative method~\cite{thomson} to Eqs.~(\ref{eq1a}) and~(\ref{eq1b}), the equations for $\langle a^{(d,s)}_{\vk} \rangle$ are obtained in the Laplace space as an expansion in powers of the coupling constant $\gamma_0^2$:
\begin{eqnarray}
\label{eq_moy1} c\,q\,\,\langle \hat{a}^{(d)}_{\vk_d}\rangle \! = \! \Big[ \frac{\gamma_0^2}{\gamma+\nu_s+\Delta\omega_s}+\mathcal{O}(\gamma_0^4) \Big]\,\langle \hat{a}^{(d)}_{\vk_d}\rangle ,\,\,\\
\label{eq_moy2} \big(\gamma + \nu_s\big) \,\,\langle\hat{a}^{(s)}_{\vk_s}\rangle \! = \! \Big[ \frac{\gamma_0^2}{c\,q+\Delta\omega_d}+\mathcal{O}(\gamma_0^4) \Big]\,\langle \hat{a}^{(s)}_{\vk_s}\rangle ,\,\,
\end{eqnarray}
where $\langle a_{\vk}^{(d,s)}\rangle=(2\pi)^{-2}\,\int dq\,d\gamma\,\langle\hat{a}^{(d,s)}_{\vk}\rangle\,{\rm e}^{q\,z+\gamma\,t}$. By introducing the half aperture angle of the pump $\theta_p = (k_0\,\rho_0)^{-1}$ and the scattering angle $\theta_d = \vert \vk_d\vert/k_0$, one obtains for the spectral widths $\Delta\omega_s/\Omega^{(s)}_{\vk_s} = \sqrt{2/\pi}\,\theta_p/\theta_d$ and $\Delta\omega_d/\omega_0 = \sqrt{2/\pi}\,\theta_p\,\theta_d$. In what follows, our attention will be focused on the case $\theta_d \gtrsim \theta_p$ and we will consider the ordering $\nu_s \ll \Delta\omega_s \ll \Delta\omega_d$. The validity of Eqs.~(\ref{eq_moy1}) and~(\ref{eq_moy2}) requires the expansion in powers of $\gamma_0^2$ to converge, which is assured when $K_B \ll 1$, where $K_B \equiv \gamma_0^2 \slash \big[(\gamma+\Delta\omega_s)\,(cq+\Delta\omega_d)\big]$ is the so-called Kubo number. 

Let us first consider the evolution of the scattered wave amplitude $\langle a^{(d)}_{\vk_d} \rangle$. The saddle point analysis of Eq.~(\ref{eq_moy1}) leads to the following behavior: for large times $t > t^{(d)}_{sat}\equiv (\gamma_0/\Delta\omega_s)^2\,z/c$, one obtains a spatial amplification: $\langle a^{(d)}_{\vk_d} \rangle \simeq \exp q^{(d)}_{incoh}\,z$, where $q^{(d)}_{incoh} \equiv \gamma_0^2/(c\,\Delta\omega_s)$. This result is similar to the one obtained in Ref.~\onlinecite{maximov_POP_01}: the spatial growth rate of the average amplitude $\langle a^{(d)}_{\vk_d}\rangle$ is limited by the pump incoherence. Here, both temporal and spatial growth rate are negligible compared to the pump spectral widths $\Delta\omega_s$ and $\Delta\omega_d$, and $K_B=\gamma_0^2/(\Delta\omega_s\,\Delta\omega_d)$. The condition for the validity of Eq.~(\ref{eq_moy1}) can then be rewritten by introducing the average power in a speckle $\langle P\rangle=\pi\,\rho_0^2\,\langle I\rangle$ and becomes $K_B \simeq \langle P \rangle/ P_c \ll 1$. For shorter times $t<t^{(d)}_{sat}$, the scattered wave behaves asymptotically in time and $z$ as: $\langle a^{(d)}_{\vk_d} \rangle \simeq \exp \big(2\,\gamma_0\,\sqrt{t\,z/c}-\Delta\omega_s\,t \big)$. In this transient regime, the spatial growth rate is smaller than $q^{(d)}_{incoh}$ and thus stays negligible compared to $\Delta\omega_d$. The condition for the validity of Eq.~(\ref{eq_moy1}) in this regime becomes $t < t^{(d)}_{val} \equiv (\Delta\omega_d/\gamma_0)^2\,z/c$, which is always fulfilled in the transient regime $t<t^{(d)}_{sat}$. Equation~(\ref{eq_moy1}) is therefore valid in both regimes under the condition $\langle P \rangle \ll P_c$. In both, convective and transient regimes, the growth of $\langle a^{(d)}_{\vk_d} \rangle$ is strongly reduced by the pump incoherence. In what follows, this result will be referred to as the incoherent amplification of FSBS.

A similar analysis of Eq.~(\ref{eq_moy2}) provides the evolution of the ion acoustic wave average amplitude $\langle a^{(s)}_{\vk_s}\rangle$. For times $t>t^{(s)}_{sat} \equiv (\gamma_0^2/\nu_s)^2\,z/c$, the ion acoustic wave grows spatially: $\langle a^{(s)}_{\vk_s} \rangle \simeq \exp (q_{coh}^{(s)}-\Delta\omega_d/c)\,z$, where $q_{coh}^{(s)}=\gamma_0^2/(c\,\nu_s)$ is the spatial growth rate of FSBS driven by a coherent pump. In this limit, $K_B = \nu_s/\Delta\omega_s$, and the condition for the validity of Eq.~(\ref{eq_moy2}) is fulfilled. For shorter times $t<t^{(s)}_{sat}$, the ion acoustic wave is in a transient regime $\langle a^{(s)}_{\vk_s} \rangle \simeq \exp\big( 2\,\gamma_0\,\sqrt{t\,z/c}-\Delta\omega_d\,z-\nu_s\,t\big)$. In this regime, Eq.~(\ref{eq_moy2}) is valid only for times $t>t^{(s)}_{val} \equiv (\gamma_0^2/\Delta\omega_s)^2\,z/c$. This time $t^{(s)}_{val} \ll t^{(s)}_{sat}$ does not exceed a few picoseconds for a millimetric plasma. Hence, the condition for the validity of Eq.~(\ref{eq_moy2}) is quickly satisfied.

The condition for spatial growth of $\langle a^{(s)}_{\vk_s} \rangle$, $c\,q_{coh}^{(s)}>\Delta\omega_d\,,$ defines a threshold for the average power in a speckle: $\langle P\rangle / P_c > \sqrt{2/\pi}\,(\nu_s/\Omega^{(s)}_{\vk_s})\,(\theta_d/\theta_p)$. Above this threshold, one observes a coherent spatial amplification $\langle a^{(s)}_{\vk_s} \rangle \simeq \exp q_{coh}^{(s)}\,z$, that is not reduced by the pump incoherence. This result is so far valid only for times $t>t^{(s)}_{sat}$. However, defining the effective spatial growth rate in the transient regime as $q_{eff}^{(d)} \equiv \big(\gamma_0^2\,c\,t/z\big)^{1/2}-\Delta\omega_d$, one obtains that the pump incoherence does not affect the spatial amplification of $\langle a^{(s)}_{\vk_s} \rangle$ for $t \gg (\Delta\omega_d/\gamma_0)^2\,z/c$. For a millimetric plasma, this time ranges between a few tenth to a few hundreds picoseconds, much shorter than the characteristic durations of ICF laser pulses.

This lead us to the following figure of merit (FOM):
\begin{eqnarray}\label{eq_critere}
C \equiv \sqrt{\frac{\pi}{2}}\,\gamma_T\,\frac{\langle P\rangle/P_c}{\nu_s/\Omega^{(s)}_{\vk_s}}\,,
\end{eqnarray}
where $\gamma_T$ has been introduced to account for the enhanced excitation of density perturbations by thermal effects. In conditions relevant for ICF, this factor can be estimated as a function of the plasma charge $Z$ and electron-ion mean-free-path $\lambda_{ei}$: $\gamma_T = 1+1.66\,Z^{5/7}(\rho_0/\lambda_{ei})^{4/7}$~\cite{brantov_POP_99}. This FOM has a straightforward physical meaning. The light scattered outside the cone with aperture $C\,\theta_p$ is amplified in the incoherent regime. On the contrary, the light scattered inside this cone demonstrates a strong, coherent amplification. Thus, for $C<1$, coherent excitation of FSBS occurs only for scattering angles $\theta_d < \theta_p$, {\it i.e.} in the cone of the incident wave. Enhanced FSBS in the incident aperture enhances plasma induced smoothing and in turn reduces the reflectivity of backward instability. Conversely for $C>1$, coherent excitation of FSBS can occur outside the incident cone. In this regime $C>1$, beam spray is thus expected. Because the ion acoustic damping rate $\nu_s/\Omega^{(s)}_{\vk_s} \ll 1$, threshold~(\ref{eq_critere}) is well below the SF threshold.


The effect of FSBS on beam spray has been confirmed in three-dimensional (3D) simulations performed with the interaction code PARAX~\cite{riazuelo_POP_00}. The laser propagation is described within the paraxial approximation. In current simulations, only the ponderomotive force is retained in the ion acoustic wave response. The Helium plasma has an electron density $0.03\,n_c$, and the electron and ion temperatures are 500 and 50 eV, respectively. The ion acoustic wave velocity is $c_s\simeq 0.17\,{\rm \mu m/ps}$ and the critical power for SF is $P_c \simeq 640\,{\rm MW}$. The ion acoustic damping rate, $\nu_s/\Omega^{(s)}_{\vk_s}=2.75$, $5.5$ or $8.25\,\%$, is chosen independently. A Gaussian laser beam with $\lambda_0=1.053\,\mu$m is focused through a random phase plate providing a speckle pattern with coherence width $\rho_0\simeq 4.3\,{\rm \mu m}$. The corresponding Rayleigh length  of the speckle is $L_R=k_0\,\rho_0^2\simeq 110\,{\rm \mu m}$. In order to avoid SF and FI, the average intensity is varied in the range $(1.1-16)\times 10^{13}\,{\rm W/cm^2}$, corresponding to variation of the average power in a speckle $\langle P\rangle =\pi\,\rho_0^2\,\langle I\rangle$ from $1$ to $14\,\%$ of $P_c$. A particular attention is paid to make sure that there is no speckle in the focal spot carrying a power above $P_c$. MS on density perturbations driven by the randomized beam, which is a non stimulated process, is kept to a rather low level by introducing a linear ramp-time $t_m\simeq 130\,{\rm ps}\gg\rho_0/c_s$ in the laser intensity profile, then decreasing the initial density perturbations in the plasma~\cite{lissage_plasma2}.

\begin{figure}
\includegraphics*[width=7.0cm]{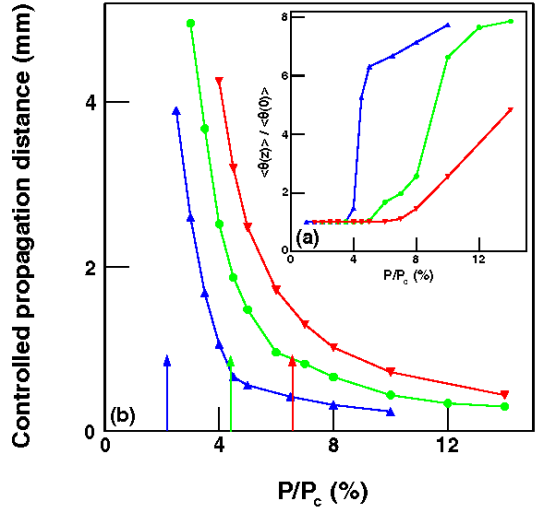}
\caption{(a) Dependence of the aperture angle of the transmitted laser beam on the normalized average power $\langle P_{sp}\rangle/P_c$ at $t=2\,$ns and $z=1.2\,$mm. (b) Propagation distance over which the aperture angle remains below $1.2\,\langle\theta(0)\rangle$. Arrows show the theoretical threshold (\ref{eq_critere}). The ion acoustic damping rates are $\nu_s/\Omega^{(s)}_{\vk_s}=2.75\,\%$ (blue), $5.5\,\%$ (green) and $8.25\,\%$ (red).}
\label{fig1}
\end{figure}

The angular aperture of the light, defined as $\langle\theta(z)\rangle = k_0^{-1}\,\langle \vk^2 \rangle^{1/2} = k_0^{-1}\, \big[ \int\! d \vk\, \vk^2\,n(\vk) \big\slash\! \int\! d \vk\,n(\vk) \big]^{1/2}$, where $n(\vk)$ is the laser spatial spectrum, has been calculated all along the propagation through the plasma. It increases with time and with $z$. The aperture angle of the light transmitted after 1.2~mm propagation is shown in Fig.~\ref{fig1}a at $t=2$~ns, as a function of the normalized average power in a speckle. Each curve corresponds to a given ion acoustic wave damping and consists of two parts. At low $\langle P\rangle/P_c$, the angular aperture of the transmitted light is equal to $\langle \theta(0) \rangle$: no angular spreading is observed. For higher powers, the aperture angle of the transmitted light increases with $\langle P\rangle/P_c$, leading to strong beam spray at high intensities.

In a complementary way, the distance over which the laser propagates with an aperture angle below $1.2\,\langle\theta(0)\rangle$ is shown as  function of $\langle P\rangle/P_c$ for different ion acoustic damping rates in Fig.~\ref{fig1}b at $t=2~$ns. At low $\langle P\rangle/P_c$, the laser propagates over a long distance (a few mm) without suffering angular spreading. For higher powers, FSBS-induced beam spray limits the propagation to a few speckle Rayleigh lengths. The transition between controlled propagation and beam spray clearly depends on the ion acoustic damping rate. A rather good agreement is found between numerical results and theoretical predictions~(\ref{eq_critere}), shown in Fig.~\ref{fig1} as vertical arrows.

\begin{figure}
\includegraphics*[width=7.0cm]{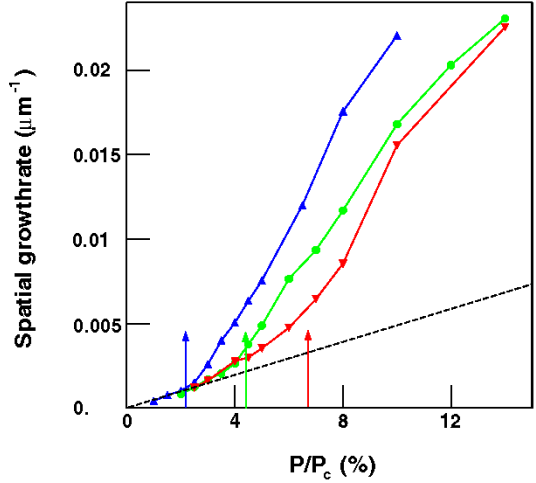}
\caption{Dependence of the spatial growth rate of the scattered wave intensity after 2~ns on the normalized average power $\langle P\rangle/P_c$. Same notations as in Fig. \ref{fig1}. The dashed curve shows the incoherent growth rate.}
\label{fig2}
\end{figure}

The temporal spectrum of the laser light has been computed taking a temporal window between 1 and 2~ns. The spatial growth rate of light scattered at the frequency $\omega_0 - 2\,c_s/\rho_0$ has been estimated. It is plotted as a function of $\langle P\rangle/P_c$ in Fig.~\ref{fig2}. In the low power regime, the spatial growth rate is less than $0.5\times10^{-2}\,{\rm \mu m^{-1}}$. It evolves almost linearly with $\langle P\rangle/P_c$ and agrees rather well with predictions in the incoherent regime (dashed line). These numerical results confirm that, below the treshold power~(\ref{eq_critere}), the growth of FSBS is limited by the pump incoherence. On the contrary, when the power in a speckle is above the threshold power (see vertical arrows), much higher growth rates, in the range from $(0.5-2.5)\,10^{-2}~{\rm \mu m^{-1}}$ are observed. These results are consistent with expectations for the coherent regime.


Our theory of FSBS driven by a spatially smoothed, monochromatic, laser beam is complementary to that developed by Lushnikov and Rose~\cite{lushnikov} for a broadband pump with small coherence time, $\tau_c \ll \rho_0/c_s$. The conclusions concerning the coherent FSBS in this regime are similar as well as the predicted threshold. Our approach allows a better understanding of the transition from controlled beam propagation to FSBS-induced beam spray. In particular, according to the present analysis, the regime referred to as equilibrium in Ref.~\onlinecite{lushnikov} may rather be an incoherent regime of FSBS with weak, albeit existing, growth. A similar approach for stimulated Raman scattering driven by an broadband pump was discussed recently in Ref.~\onlinecite{santos_PRL_07}. In connection with the results of Refs.~\cite{lushnikov} and~\cite{santos_PRL_07}, we may conclude that laser beam incoherence can not prevent coherent growth whenever the temporal or the spatial growth rate of any of the daughter waves is larger that the corresponding  effective bandwidth due to the pump incoherence.

Beam spray has been reported recently in experiment~\cite{froula_POP_07}. The authors observe a threshold for the angular beam spreading and attribute it to FI. For experimental conditions ($n_0=0.06\,n_c$, $T_e=3.5\,{\rm keV}$, $\lambda_0=0.351\,{\rm \mu m}$, $\nu_s/\Omega^{(s)}_{\vk_s}\simeq 0.15$ and the f-number $f_{\sharp}=6.7$), the FOM~(\ref{eq_critere}) predicts beam spray for average power in a speckle $\langle P \rangle \simeq 7.5\,\%$ of $P_c$, where we have estimated $Z \simeq 5.3$ for the CH plasma, $\lambda_{ei}\simeq 75\,{\rm \mu m}$ and $\rho_0 \simeq (2/\pi)\,\lambda_0\,f_{\sharp}$. It is in perfect agreement with the average power at the measured intensity threshold $\langle I\rangle \simeq 2\times 10^{15}\,{\rm W/cm^2}$: $\langle P\rangle \simeq \pi\,\rho_0^2\,\langle I\rangle\simeq 0.07\,P_c$. It is significantly below the SF threshold, which prompts us to suggest that the observed beam spray is a direct consequence of the coherent FSBS rather than SF or FI. Last but not least, the dependence of the threshold~(\ref{eq_critere}) on the ion acoustic damping provides an important insight on the control of propagation of randomized laser beam through plasmas. We propose from Eq.~(\ref{eq_critere}) a new, theoretically derived, criterion for beam spray:
\begin{eqnarray}
0.1\,\gamma_T\,\frac{\Omega^{(s)}_{\vk_s}}{\nu_s}\,\lambda_{0}^2{\rm [\mu m]}\,I_{13}\,\frac{n_0}{n_c}\,\frac{3}{T_e\,{\rm [keV]}}\,\left(\frac{f_{\sharp}}{8}\right)^2>1\,,
\end{eqnarray}
where $I_{13}$ is the laser average intensity in $10^{13}\,{\rm W/cm^2}$ . Taking $\nu_s/\Omega^{(s)}_{\vk_s} \simeq 0.15$ and $\gamma_T \simeq 1.6$ corresponding to the experimental conditions, this criterion leads exactly to the FOM suggested in Eq.~(4) of Ref.~\onlinecite{froula_POP_07}.


In conclusion, a statistical model for FSBS driven by an angularly broadened pump has been developped. It predicts a fast (coherent) growth of the daughter waves well below the SF threshold. This instability leads to a strong deterioration of the beam propagation which is not stabilized by the laser beam spatial smoothing. A threshold condition is derived providing a new FOM for the control of beam propagation through plasmas. The theoretical model is confirmed by 3D numerical simulations and provides an explanation for a recent experiment~\cite{froula_POP_07}. This transition from incoherent to coherent growth of FSBS associated to beam spray is of crucial importance for the design of forthcoming ICF experiments.

Financial support from ANR under contract CORPARIN is acknowledged.

\newpage$\,$

\newpage$\,$


\begin{thebibliography}{2}
\bibitem{remark} Time derivatives in Eq.~(\ref{eq1a}) has been neglected, which requires the time for the laser light to travel through a speckle $L_R/c$ to be much shorter than both, the laser coherence time and the characteristic time for the acoustic time to cross a speckle $\rho_0/c_s$. These conditions are fulfilled in most situations of interest.
\end{thebibliography}
\end{document}